\documentclass[sigconf]{acmart}
\AtBeginDocument{%
  }

\copyrightyear{2026}
\acmYear{2026}
\setcopyright{cc}
\setcctype{by}
\acmConference[WWW '26]{Proceedings of the ACM Web Conference 2026}{April 13--17, 2026}{Dubai, United Arab Emirates}
\acmBooktitle{Proceedings of the ACM Web Conference 2026 (WWW '26), April 13--17, 2026, Dubai, United Arab Emirates}
\acmPrice{}
\acmDOI{10.1145/3774904.3792825}
\acmISBN{979-8-4007-2307-0/2026/04}

\renewcommand\footnotetextcopyrightpermission[1]{}  

\usepackage{microtype}
\usepackage{balance}
\usepackage{xcolor}
\usepackage[table]{xcolor}
\usepackage{graphicx}
\usepackage{svg}
\usepackage{subcaption}
\usepackage{pifont}
\usepackage{marvosym}
\usepackage{enumitem}
\setlist[itemize]{left=0pt}
\usepackage{multirow}
\usepackage{booktabs}
\usepackage{threeparttable}
\usepackage{etoolbox}
\begin{document}

\title{Not All Candidates are Created Equal: A Heterogeneity-Aware Approach to Pre-ranking in Recommender Systems}

\settopmatter{printacmref=false}
\settopmatter{authorsperrow=4}
 \emergencystretch=3em 
\author{Pengfei Tong}
\affiliation{%
  \institution{ByteDance}
   \city{Hangzhou}
   \country{P.R.China}
}
\email{tongpengfei@bytedance.com}

\author{Siyuan Chen}
\affiliation{%
  \institution{ByteDance}
   \city{Beijing}
   \country{P.R.China}
}
\email{chensiyuan.279}
\email{@bytedance.com}

\author{Chenwei Zhang}
\affiliation{%
  \institution{ByteDance}
   \city{Shenzhen}
   \country{P.R.China}
}
\email{zhangchenwei.929}
\email{@bytedance.com}

\author{Bo Wang}
\affiliation{%
  \institution{ByteDance}
   \city{Beijing}
   \country{P.R.China}
}
\email{wangbo.9830@bytedance.com}

\author{Qi Pi \textsuperscript{\dag} }
\thanks{\dag Corresponding author.}
\affiliation{%
  \institution{ByteDance}
   \city{Hangzhou}
   \country{P.R.China}
}
\email{wk@bytedance.com}

\author{Pixun Li}
\affiliation{%
  \institution{ByteDance}
   \city{Beijing}
   \country{P.R.China}
}
\email{lipixun@bytedance.com}

\author{Zuotao Liu}
\affiliation{%
  \institution{ByteDance}
   \city{Shanghai}
   \country{P.R.China}
}
\email{michael.liu@bytedance.com}

\renewcommand{\shortauthors}{Pengfei Tong et al.}

\begin{abstract}
Most large-scale recommender systems follow a multi-stage cascade of retrieval, pre-ranking, ranking, and re-ranking.
A key challenge at the pre-ranking stage arises from the heterogeneity of training instances sampled from coarse-grained retrieval results, fine-grained ranking signals, and exposure feedback.
Our analysis reveals that prevailing pre-ranking methods, which indiscriminately mix heterogeneous samples, suffer from gradient conflicts: hard samples dominate training while easy ones remain underutilized, leading to suboptimal performance.
We further show that the common practice of uniformly scaling model complexity across all samples is inefficient, as it overspends computation on easy cases and slows training without proportional gains.
To address these limitations, this paper presents \textbf{H}eterogeneity-Aware \textbf{A}daptive \textbf{P}re-ranking (\textbf{HAP}), a unified framework that mitigates gradient conflicts through conflict-sensitive sampling coupled with tailored loss design, while adaptively allocating computational budgets across candidates.
Specifically, HAP disentangles easy and hard samples, directing each subset along dedicated optimization paths.
Building on this separation, it first applies lightweight models to all candidates for efficient coverage, and further engages stronger models on the hard ones, maintaining accuracy while reducing cost.
This approach not only improves pre-ranking effectiveness but also provides a practical perspective on scaling strategies in industrial recommender systems.
HAP has been deployed in the Toutiao production system for 9 months, yielding up to \textbf{0.4\%} improvement in user app usage duration and \textbf{0.05\%} in active days, without additional computational cost.
We also release a large-scale industrial hybrid-sample dataset to enable the systematic study of source-driven candidate heterogeneity in pre-ranking.
\end{abstract}



\maketitle

\vspace{-0.2cm}
\section{Introduction}

Modern recommender systems aim to identify content that matches user interests from hundreds of millions of candidates.
To balance effectiveness and efficiency, multi-stage cascades have emerged as the prevailing paradigm \cite{covington2016deep, liu2017cascade, gao2024causal, qin2022rankflow}, comprising retrieval \cite{huang2013learning, zhu2018learning, zhou2021contrastive}, pre-ranking \cite{ma2021towards, song2022rethinking, wang2020cold}, ranking \cite{qi2020searchbasedusermodelinglifelong, pi2019practice, zhou2020can}, and re-ranking \cite{lin2024discrete, liu2022neural, wang2019sequential} stages.
Positioned between retrieval and ranking, pre-ranking filters thousands of retrieved items down to a few hundred for subsequent ranking within milliseconds.
A central challenge here lies in the diversity of the candidates: while many items are trivially irrelevant and easy to filter, others are near-positives that closely resemble ground-truth and are difficult to distinguish.
This mixture of easy and hard samples introduces inherent heterogeneity, leading to competing learning signals that challenge effective optimization.

\begin{figure}[t]
  \centering
  \includegraphics[width=1\linewidth]{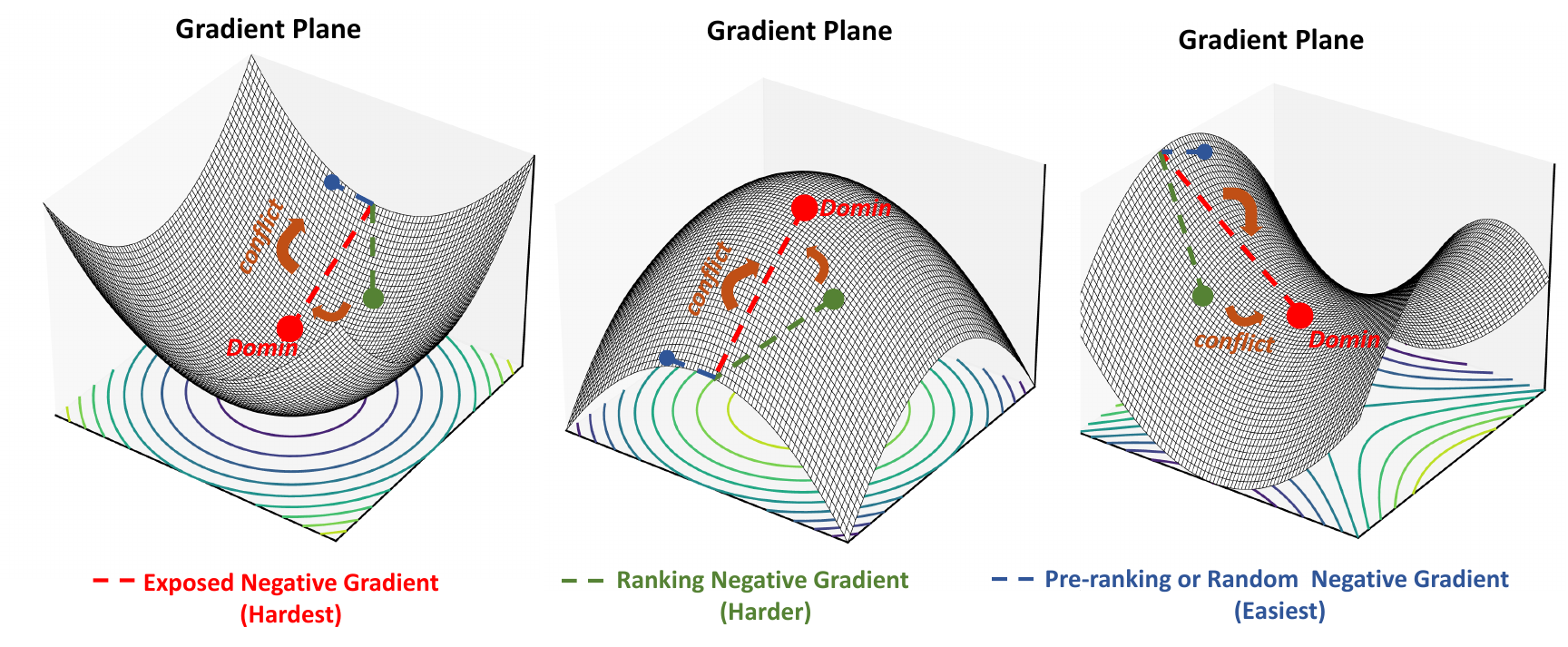}
  \caption{Gradient conflicts across negatives of varying difficulty, illustrating the inherent heterogeneity of candidates.}
  \label{fig:Gradient conflicts}
  \vspace{-15pt}
\end{figure}

Previous research has sought to address this heterogeneity, primarily through two divergent paradigms, both of which we identify as suboptimal. 
\textbf{1) Indiscriminate Data Expansion and its Optimization Pitfalls.}
A common strategy to address distributional diversity is expanding the negative training set.
Early works \cite{song2022rethinking, zhang2023rethinking} formulate the problem as Sample Selection Bias (SSB), employing hard negative mining to improve generalization across distribution shifts.
More recent work \cite{Zhao_2025hccp} has extended this idea by incorporating unexposed negatives from upstream retrieval and low-ranked items from pre-ranking and downstream ranking.
Although these strategies undoubtedly broaden data coverage, they introduce a severe optimization conflict.
The resulting mixture of easy and hard negatives causes hard samples to dominate training, steering the model toward suboptimal local minima that overfit to noise and sabotage the optimization process (Fig.\ref{fig:Gradient conflicts}).
\textbf{2) Blind Model Scaling and its Asymmetric Returns}.
Beyond data expansion, another common industrial practice is to scale up model capacity in the hope of capturing complex user–item interaction patterns.
Larger models aligned with the ranking stage have been shown to improve pre-ranking accuracy \cite{wang2020cold, zhao2023both, zhao2023copr, xu2020privileged}.
However, gains are largely confined to hard-to-rank candidates, yielding only minimal improvements on the majority of easier ones. Thus, pursuing model expressiveness without addressing gradient conflicts from heterogeneous candidates represents a suboptimal and costly strategy.

Building on these insights, we propose \textbf{H}eterogeneity-Aware \textbf{A}daptive \textbf{P}re-ranking (\textbf{HAP}), a unified framework designed to address two key challenges posed by heterogeneous candidates: optimization conflicts and computational inefficiency.
\textbf{1) Gradient-Harmonized Contrastive Learning (GHCL):} Our theoretical analysis of Binary Cross-Entropy (BCE) and InfoNCE shows that hard negatives induce disproportionately large gradients, destabilizing training.
GHCL constructs four difficulty-aware negative sets in the TouTiao\footnote{One of the largest news recommendation platforms in China\label{fn:toutiao}} platform, following prior strategies \cite{Zhao_2025hccp}, and introduces a tailored contrastive loss that harmonizes gradient contributions across them.
This approach balances easy and hard samples, ensures comprehensive distributional coverage, and prevents dominance of hard negatives, leading to stable convergence and stronger generalization.
\textbf{2) Difficulty-Aware Model Routing (DAMR):} Not all candidates require the same modeling complexity.
DAMR employs a progressive routing strategy.
A lightweight model first processes all candidates for broad coverage, while difficult samples are selectively forwarded to a more expressive model, allocating computation efficiently within a fixed budget.

Through extensive qualitative and quantitative experiments, we demonstrate that \textbf{HAP} effectively addresses the challenges from heterogeneous candidate distributions and significantly improves pre-ranking performance without additional engineering costs.
Our contributions are summarized as follows:
\begin{itemize}
\item \textbf{Heterogeneity-Aware Pre-Ranking Framework with Theoretical Foundations.} We propose a unified framework that tackles massive, heterogeneous candidates in pre-ranking.
Grounded in a theoretical and empirical study of sample construction and gradient conflicts, \textbf{HAP} achieves an effective trade-off between recommendation effectiveness and computational efficiency.
\item \textbf{Open Multi-Stage Dataset.} We release a large-scale, fully annotated multi-stage recommendation dataset in TouTiao\footref{fn:toutiao} to facilitate reproducible research and stimulate future advancements in pre-ranking and multi-stage recommendation.
\item \textbf{Notable Practical Gains.} We deployed this framework in our industrial-scale recommender system, Toutiao\footref{fn:toutiao}, and achieved a 0.05\% increase in user active days and a 0.4\% improvement in app usage duration, representing substantial gains at this scale.
\end{itemize}

\section{RELATED WORK}
Research on pre-ranking mainly advances along two directions: sample selection optimization and model architecture refinement.

\textbf{Sample section optimization} aims to mitigate exposure bias by broadening the training distribution, for instance, by incorporating unexposed or low-ranked items \cite{ma2020off, 10.1145/3564284, Gao_Zhou_Shao_Wu_Gao_Ren_Qi_Deng_Liu_2025}.
Recent studies adopt contrastive learning and counterfactual modeling \cite{zhang2023rethinking,qin2022rankflow} to enhance generalization toward unseen candidates.
IRRD \cite{10.1016/j.ins.2021.08.060} introduces item-side regularization to align ranking behaviors, while HCCP \cite{Zhao_2025hccp} explores cross-stage coordination to improve generalization on low-ranked items.
However, these methods treat all candidates uniformly and ignore the inherent distributional heterogeneity, which can cause gradient interference during optimization.

\textbf{Model architecture refinement} focuses on improving representation capacity.
Deep neural networks (DNNs) have become standard for dot-product-based matching, independently embedding users and items to compute relevance scores \cite{huang2013learning, palangi2014semantic}.
Subsequent research \cite{wang2020cold, ma2021towards, li2022autofas, li2022inttower} investigates advanced architectures and feature modeling techniques to enhance model's ability.
COLD \cite{wang2020cold} employs feature selection to capture salient signals, FSCD \cite{ma2021towards} performs adaptive feature pruning via variational dropout, and COPR \cite{zhao2023copr} aligns pre-ranking with fine-stage outputs.
While such methods strengthen modeling for hard samples, they bring limited gains for easy ones, resulting in asymmetric returns and increased computational costs.
Moreover, scaling model size does not resolve the underlying gradient conflicts caused by heterogeneous sample difficulty, rendering these strategies both inefficient and insufficient.

In summary, existing efforts have advanced pre-ranking through better expressiveness, broader sample coverage, and stage coordination, yet they overlook the dual challenge of heterogeneous sample optimization and efficient model allocation.
To overcome these limitations, we introduce HAP, a heterogeneity-aware pre-ranking framework that jointly harmonizes gradient learning and adaptively routes candidates based on their difficulty, achieving both effective optimization and efficient computation.
\section{METHODOLOGY}

\subsection{Preliminaries}
\begin{figure*}[t]
  \centering
  \includegraphics[width=0.88\linewidth]{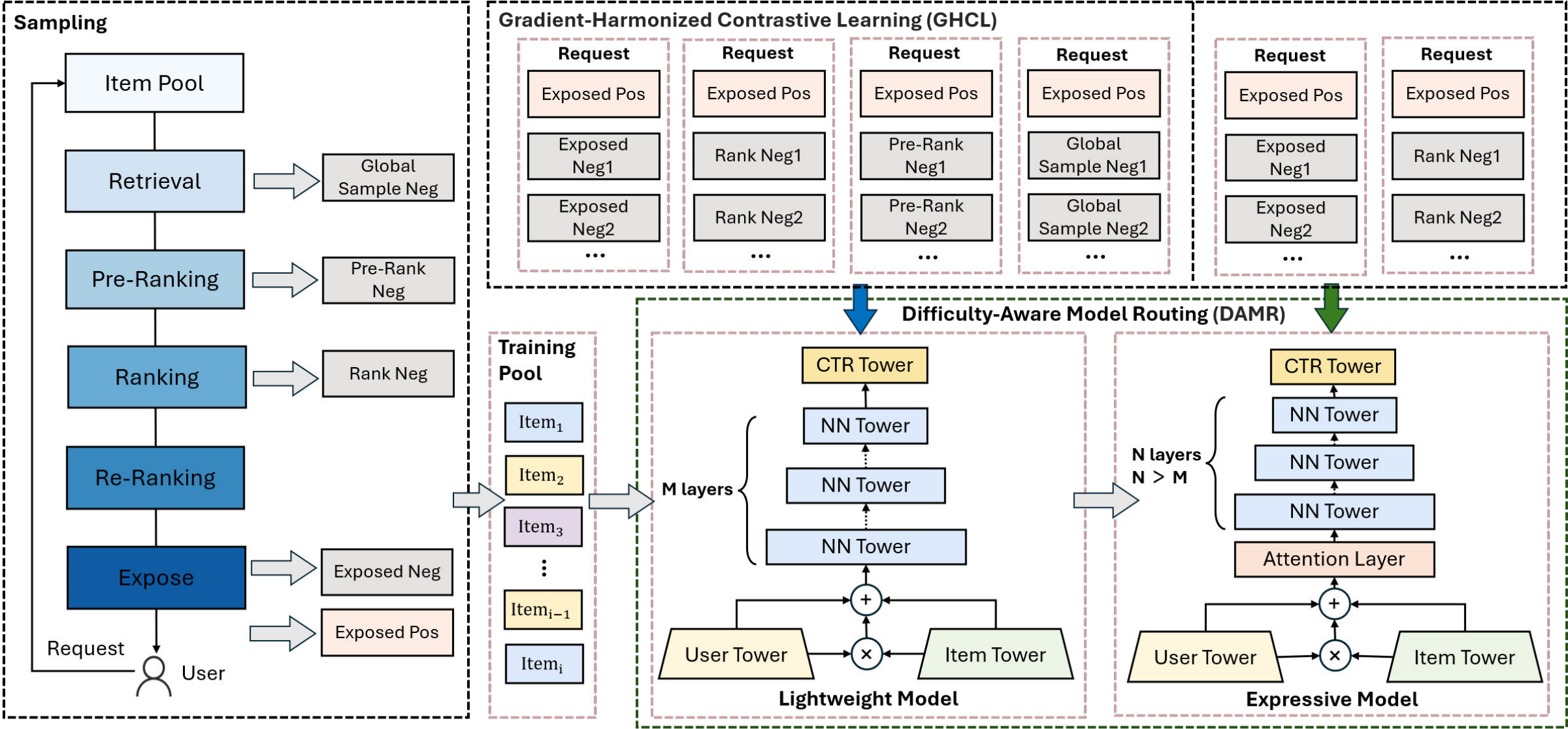}
  \caption{Architecture of HAP’s training sample collection and model training. Samples are drawn from complete requests, including exposed positives and negatives, and negatives from retrieval, pre-ranking , and ranking stages. HAP adopts the DAMR architecture: the lightweight model learns from all sample types via GHCL, scores retrieval candidates, and filters easy ones, while the expressive model trains exclusively on hard samples using a deeper, attention-based network.}
  \label{fig:framework}
  \vspace{-10pt}
\end{figure*}

We begin by detailing the data composition in the pre-ranking recommendation setting.
For each user request $q$, the system processes billions of candidates through a multi-stage pipeline and returns a set of top-ranked items $E^q=\{e_1, e_2, \ldots , e_n\}$ for final exposure, conditioned on user information $u$, where $e_i$ donates an exposed content and $n$ is the maximum number of items per session.
At stage $k$, the candidate set is represented as $\mathcal{C}_k^q = \{ c_1, c_2, \ldots,  c_l \}$, where $c_l$ is the $l$-th candidate at that stage.
Training samples derived from a single request are categorized into five types:
\begin{itemize}
\item \textbf{Exposed positives (EP):} Items that user actually clicked, defined as $\mathcal{S}_{\text{EP}}=\{(u, x_i)| x_i \in E^q, \text{click}=1\}$.
These samples capture verified user interests.
\item \textbf{Exposed negatives (EN):} Items that were displayed but not clicked, defined as $\mathcal{S}_{\text{EN}}=\{(u, x_i)| x_i \in E^q, \text{click}=0\}$.
These provide explicit negative feedback signals.
\item \textbf{Ranking negatives (RN):} Candidates ranked below a threshold $\text{Rank}_\text{L}$ in the subsequent ranking stage, defined as $\mathcal{S}_{\text{RN}}=\{(u, x_i)| x_i \in \mathcal{C}_{\text{ranking}}^q, i>\text{Rank}_\text{L} \}$. These samples help align pre-ranking with downstream ranking objectives.
\item \textbf{Pre-ranking negatives (PRN):} Candidates scored below a threshold $\text{Prerank}_\text{L}$ by the current pre-ranking stage, defined  as $\mathcal{S}_{\text{PRN}}=\{(u, x_i)| x_i \in \mathcal{C}_{\text{pre-ranking}}^q, i>\text{Prerank}_\text{L} \}$.
These enhance the model’s ranking discrimination by widening the score gap between top- and low-ranked candidates.
\item \textbf{Global random negatives (GN):} Randomly sampled candidates from retrieval, defined as~$\mathcal{S}_{\text{GN}}= \{x_i \sim \text{Uniform}(\mathcal{C}_{\text{retrieval}}^q)\}_1^N$, where $N$ is the sample size.
These mitigate sample selection bias (SSB) and improve distributional coverage.
\end{itemize}
Although diverse samples enhance model generalization, they also introduce substantial heterogeneity in data distribution and learning difficulty.
Such heterogeneity presents a key challenge for unified optimization, motivating our heterogeneity-aware design.
The overall framework is depicted in Fig.~\ref{fig:framework}.

\subsection{Gradient Dominance from Heterogeneity}
To understand how sample heterogeneity affects optimization, we analyze the gradient behavior of different sample types under commonly used pre-ranking loss functions.
We categorize samples by their learning difficulty and examine their gradient norms (\textit{Grad Norm}), logit values (\textit{Logits}), and gradient cosine similarities (\textit{Cosine}), as summarized in Table~\ref{tab:gradient_analysis_comparison}.
Our analysis reveals that logit values generally increase with sample hardness.
Formally, for a user \( u \), the logits for hard negatives and easy negatives are as follows:
\begin{equation}
z_u^{\text{hn}} = f(x_u^i, \theta),\quad x_u^i \in D_u^{\text{hn}}
\end{equation}
\begin{equation}
z_u^{\text{en}} = f(x_u^i, \theta),\quad x_u^i \in D_u^{\text{en}}
\end{equation}
Empirically, we observe \( z^{\text{hn}}_u > z^{\text{en}}_u \), which is intuitive since hard negatives are semantically closer to positives and thus receive higher model scores.
We further find that hard negatives induce larger gradient norms and exhibit lower cosine similarity with easier samples, indicating conflicting gradient directions.
To explain these observations, we analyze the gradient formulations of the Binary Cross-Entropy (BCE) and InfoNCE losses—two representative objectives for pre-ranking models.

\begin{table}[t]
\centering
\caption{Comparison of gradient characteristics across negatives of varying difficulty under BCE and InfoNCE losses. Sample types: exposed (EN), ranking (RN), pre-ranking (PRN), and global random (GN).}
\label{tab:gradient_analysis_comparison}
\small
\begin{tabular}{@{}llcccc@{}}
\toprule
\textbf{Loss} & \textbf{Sample} & \textbf{Difficulty} & \textbf{Grad Norm} & \textbf{Logits} & \textbf{Cosine} \\
\midrule
\multirow{4}{*}{\textbf{BCE}} & EN & Hardest & 1.101 & 0.322 & -- \\
& RN & Harder & 0.254 & 0.232 & 0.816 \\
& PRN & Easier & 0.140 & 0.191 & 0.496 \\
& GN & Easiest & 0.060 & 0.173 & 0.237 \\
\midrule
\multirow{4}{*}{\textbf{InfoNCE}} & EN & Hardest & 1.750 & 0.326 & -- \\
& RN & Harder & 0.401 & 0.235 & 0.820 \\
& PRN & Easier & 0.352 & 0.192 & 0.607 \\
& GN & Easiest & 0.162 & 0.176 & 0.576 \\
\bottomrule
\end{tabular}
\vspace{-10pt}
\end{table}
\subsubsection{Gradient Dominance in BCE}
BCE loss is widely adopted for recommendation tasks formulated as binary classification.
For a user request \( q \), the training data is defined as $\mathcal{D}^\text{BCE}_q = \{(u_q, x_i, y_i)\}_{i=1}^{N_q}$, where $N_q$ is the number of samples, and each tuple represents a user-item pair with a binary label \( y_i \).
A clicked item is labeled \( y_i = 1 \), while an unclicked one \( y_i = 0 \).
Let $z_i=f(u_q,x_i;\theta) \in \mathbb{R}$ denote the logit output for sample \( i \), and $\hat{y}_i=\sigma(z_i)$ be the predicted probability , where \( \sigma \) is the sigmoid function.
The BCE loss and its gradient with respect to \( z_i \) are:
\begin{equation}
\mathcal{L}_{\text{BCE}} = \sum_{q} \sum_{i=1}^{N_q} \left[ -y_i \log \sigma(z_i) - (1 - y_i) \log (1 - \sigma(z_i)) \right]
\end{equation}
For negative samples (\( y_i = 0 \)), the gradient simplifies to \( \frac{\partial \mathcal{L}_{\text{BCE}}}{\partial z_i} = \sigma(z_i) \).
Since the sigmoid function is monotonically increasing, negatives with larger logits, which indicate higher difficulty, tend to receive stronger gradients.
As shown in Table~\ref{tab:gradient_analysis_comparison}, this results in a clear hierarchy (\( z_{\mathrm{EN}} > z_{\mathrm{RN}} > z_{\mathrm{PRN}} > z_{\mathrm{GN}} \)), where the hardest negatives dominate the gradient updates. 
Such dominance can overshadow easier samples, impeding balanced optimization and leading to unstable convergence.

\subsubsection{Gradient Dominance in InfoNCE}
The goal of the pre-ranking stage is to select the top-K candidates, making the contrastive InfoNCE loss a natural choice. 
For a user request \( q \), the data is structured as $\mathcal{D}^\text{InfoNCE}_q = \{(u_q, x_i^+,x_1^- \cdots ,x_M^-)\}_{i=1}^{N_p}$, where each tuple contains a positive item \( x^+ \) and different type negative items ${x_1^- \cdots ,x_M^-}$, and \( M \) denotes the number of negative samples and \( N_p \) denotes the number of positive samples.
The model $f_{\theta}$ computes the similarity between user $u$ and an item $s^+ = f(x_i^+,u_q;\theta)$ for positives and $s_j^- = f(x_j^-,u_q;\theta)$ for negatives. For convenience, the analysis below will focuses on the InfoNCE of a single sample pair.
The InfoNCE loss and gradient for a specific negative \( s_k^- \) are:
\begin{align}
\mathcal{L}_{\text{InfoNCE}} &= -\log \frac{e^{s^+}}{e^{s^+} + \sum_j e^{s_j^-}}, \\
\frac{\partial L}{\partial s^-_k} &= \frac{e^{s_k^-}}{e^{s^+} + \sum_j e^{s_j^-}}.
\end{align}
This formulation reveals a gradient coupling effect arising from the distributional shift in negative sample difficulties.
Each negative's gradient corresponds to its respective softmax probability \( P(k^-) \); thus, extremely hard negatives (with high \( s_k^- \)) dominate the denominator, suppressing gradient contributions from all other samples.
This coupling results in unstable and biased optimization, as model updates are disproportionately influenced by the hardest samples in each batch.
Consequently, sample heterogeneity amplifies gradient magnitude imbalances and introduces inter-sample interference, hindering convergence stability.

\subsection{Gradient-Harmonized Contrastive Learning}
Our previous analysis reveals a fundamental challenge in pre-ranking: gradient conflicts and coupling among negative samples of varying difficulty.
While pointwise losses like BCE can mitigate this via simple sample reweighting, list-wise losses such as InfoNCE are more sensitive due to gradient interdependence induced by softmax normalization.
To address this, we propose a Gradient-Harmonized Contrastive Loss (GHCL), a simple yet effective modification to the origin InfoNCE objective that explicitly decouples gradients across negatives of different difficulty levels.
Formally, our GHCL is:
\begin{equation}\mathcal{L}_{\text{GHCL}} = 
\underbrace{\mathcal{L}_{\text{RN}} + \mathcal{L}_{\text{EN}}}_{\mathcal{L}_{\text{hard}}} + \underbrace{\mathcal{L}_{\text{GN}} + \mathcal{L}_{\text{PRN}}}_{\mathcal{L}_{\text{easy}}}
\end{equation}
\begin{align}
\mathcal{L}_{\text{hard}} &= -\log \frac{e^{s_p}}{e^{s_p} + \sum_{j \in \mathcal{N}_{\text{hard}}} e^{s_j^-}}, \\
\mathcal{L}_{\text{easy}} &= -\log \frac{e^{s_p}}{e^{s_p} + \sum_{j \in \mathcal{N}_{\text{easy}}} e^{s_j^-}}.
\end{align}
Here, \(s_p\) denotes the positive sample's score, and \(s_j^-\) represents the score of a negative sample from the hard set \(\mathcal{N}_{\text{hard}} \in \{ \mathcal{N}_{\text{EN}}, \mathcal{N}_{\text{RN}} \}\) and the easy set \(\mathcal{N}_{\text{easy}} \in \{ \mathcal{N}_{\text{PRN}}, \mathcal{N}_{\text{GN}} \}\), respectively.
By splitting the negatives, we effectively isolate gradient computation within each subgroup, thereby mitigating cross-group coupling.
Let $s_{\text{hk}}^-$ and $s_{\text{ek}}^-$ be scores of a hard and easy negative.
Under the original InfoNCE ($\mathcal{L}_{\text{ORG}}$), their gradients are:
\begin{align}
G^{\text{hk}}_{\text{ORG}} &= \frac{\partial \mathcal{L}_{\text{ORG}}}{\partial s_{\text{hk}}^-} = \frac{e^{s_{\text{hk}}^-}}{e^{s_p} + \sum_{j} e^{s_j^-}}, \\
G^{\text{ek}}_{\text{ORG}} &= \frac{\partial \mathcal{L}_{\text{ORG}}}{\partial s_{\text{ek}}^-} = \frac{e^{s_{\text{ek}}^-}}{e^{s_p} + \sum_{j} e^{s_j^-}}.
\end{align}
The ratio between their gradient magnitudes can thus expressed as:
\begin{equation}
R_{\text{ORG}} = \frac{G^{\mathrm{hk}}_{\text{ORG}}}{G^{\text{ek}}_{\text{ORG}}} = e^{s_{\text{hk}}^- - s_{\text{ek}}^-}.
\end{equation}
Since hard negatives typically receive higher scores ($s_{\text{hk}}^-$ > $s_{\text{ek}}^-$), this ratio can be exponentially large, indicating that hard negatives dominate the gradient and suppress learning from easier ones.

Under GHCL, the gradients are computed within each subgroup:
\begin{align}
G^{\mathrm{hk}}_{\text{GHCL}} &= \frac{e^{s_{\mathrm{hk}}^-}}{e^{s_p} + \sum_{j \in \mathcal{N}_{\text{hard}}} e^{s_j^-}}, \\
G^{\mathrm{ek}}_{\text{GHCL}} &= \frac{e^{s_{\mathrm{ek}}^-}}{e^{s_p} + \sum_{j \in \mathcal{N}_{\text{easy}}} e^{s_j^-}}.
\end{align}
The resulting gradient ratio becomes:
\begin{align}
R_{\text{GHCL}} &= \frac{G^{\mathrm{hk}}_{\text{GHCL}}}{G^{\mathrm{ek}}_{\text{GHCL}}} = e^{s_\text{hk}^- - s_\text{ek}^-} \cdot \left( \frac{e^{s_p} + \sum_{j \in \mathcal{N}_{\text{easy}}} e^{s_j^-}}{e^{s_p} + \sum_{j \in \mathcal{N}_{\text{hard}}} e^{s_j^-}} \right).
\end{align}
The term in parentheses acts as a correction factor:
\begin{equation}
C = \frac{e^{s_p} + \sum_{j \in \mathcal{N}_{\text{easy}}} e^{s_j^-}}{e^{s_p} + \sum_{j \in \mathcal{N}_{\text{hard}}} e^{s_j^-}} < 1.
\end{equation}
Therefore, the overall ratio satisfies:
\begin{equation}
R_{\text{GHCL}} = R_{\text{ORG}} \cdot C < R_{\text{ORG}}.
\end{equation}
GHCL bridges the gradient magnitude gap between hard and easy negatives, effectively harmonizing their contributions.
By balancing optimization dynamics across difficulty levels, GHCL stabilizes training and enhances representation learning, particularly in pre-ranking scenarios with varied negative sample difficulties.

\subsection{Difficulty-Aware Model Routing}
\begin{figure}[t]
  \centering\includegraphics[width=0.9\linewidth]{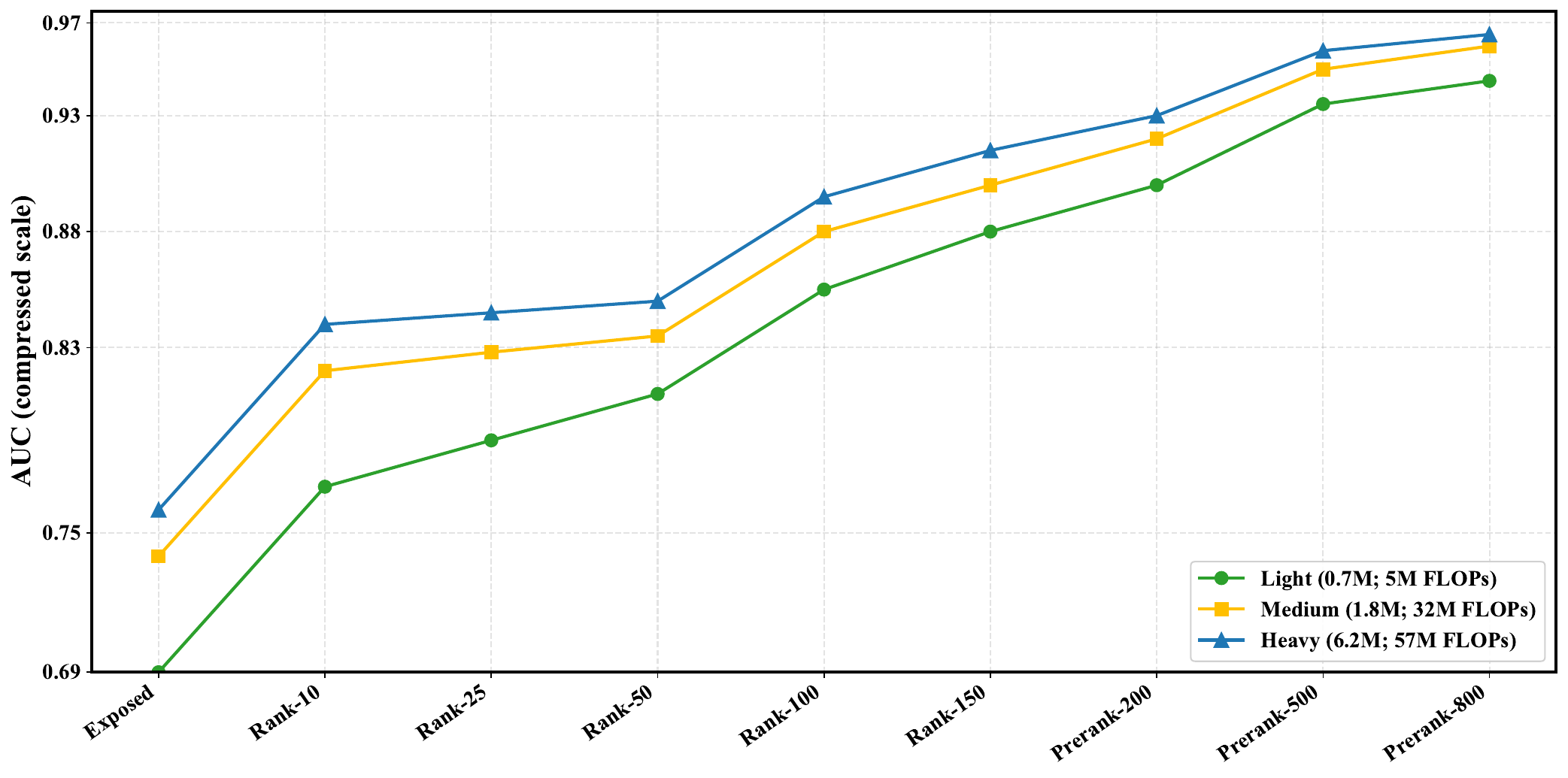}
\caption{Model AUC \emph{vs.} sample difficulty, with "Exposed" for exposed negatives and "Rank-\(N\)"/ "Prerank-\(N\)" for negatives ranked after position \(N\); smaller \(N\) implies higher difficulty.
}
  \label{fig:sample_difficult}
  \vspace{-0.5cm}
\end{figure}

While GHCL harmonizes gradient signals, it does not address the architectural inefficiency in handling heterogeneous sample difficulties.
Our empirical analysis reveals that model capacity should align with sample difficulty.
As shown in Fig.\ref{fig:sample_difficult}, lightweight models perform comparably to complex ones on easy samples but degrade sharply on hard ones.
This observation motivates a difficulty-adaptive model hierarchy, where simple models handle easy cases and expressive models specialize in hard cases.

\subsubsection{Coarse-Grained Filtering with a Lightweight Model}
A lightweight model \( f_l \) efficiently processes the full candidate set from retrieval stage, filtering out easy negatives at low computational cost.
To ensure \( f_l \) generalized across the difficulty spectrum, we train it using a diverse mixture of negatives: recall-level (GN), pre-ranking (PRN), ranking (RN), and exposed negatives (EN).
Following GHCL, we compute an independent InfoNCE losses for each negative type $t \in \{\text{EN}, \text{RN}, \text{PRN}, \text{GN}\}$:
\begin{equation}
\mathcal{L}_t = -\log \frac{e^{f(\mathbf{x}^+, \mathbf{u})}}{e^{f(\mathbf{x}^+, \mathbf{u})} + \sum_{j=1}^{n_t} e^{f(\mathbf{x}_{t,j}^-, \mathbf{u})}} ,
\label{eq:infonce_i}
\end{equation}
where $n_t$ is the number of negative samples of type $t$.
The overall loss for this stage is:
\begin{equation}
\mathcal{L}_{\text{light}} = \lambda_{\text{RN}}\mathcal{L}_{\text{RN}} + \lambda_{\text{EN}}\mathcal{L}_{\text{EN}} + \lambda_{\text{GN}}\mathcal{L}_{\text{GN}} + \lambda_{\text{PRN}}\mathcal{L}_{\text{PRN}}
\end{equation}
This stratified objective prevents overfitting to any single difficulty level and encourages generalizable features for early-stage filtering.

\subsubsection{Fine-Grained Ranking with a Complex Model}
A high-capacity model \( f_c \) performs a fine-grained ranking on the reduced candidate set from the previous stage.
Since there candidates are predominantly hard samples, \( f_c \) is specialized by training only on the most challenging negatives: exposed (EN) and ranking (RN).
The loss follows the same InfoNCE form as Eq. \ref{eq:infonce_i}:
\begin{equation}
\mathcal{L}_{\text{complex}} = \lambda_{\text{RN}}\mathcal{L}_{\text{RN}} + \lambda_{\text{EN}}\mathcal{L}_{\text{EN}}
\end{equation}
This two-stage design ensures efficient resource allocation: easy samples are processed cheaply by \( f_l \), while the expensive capacity of \( f_c \) is reserved for the hard cases that require it.
This directly resolves the architectural inefficiency identified at the outset, achieving an optimal balance of accuracy and system performance.

\subsubsection{Final Loss}
InfoNCE requires at least one positive per request, so sessions without clicks are discarded, causing significant data loss.
To address this, we introduce an auxiliary BCE loss that utilizes all samples within a batch:
\begin{equation}
\mathcal{L}_{\text{global}} = \frac{1}{|\mathcal{B}|} \sum_{(x, y) \in \mathcal{B}} \left[ - y \log \sigma(s(x)) - (1 - y) \log (1 - \sigma(s(x))) \right]
\end{equation}
Here, $\mathcal{B}$ is the set of all samples in the current batch. $y \in \{0,1\}$ is the label, and $\sigma(\cdot)$ is the Sigmoid function.
We jointly optimize the three losses mentioned above.
The final training objective is:
\begin{equation}
\mathcal{L}_{\text{total}} = \mathcal{L}_{\text{light}} + \mathcal{L}_{\text{complex}} + \alpha \cdot \mathcal{L}_{\text{global}}
\end{equation}
This unified optimization integrates hierarchical difficulty-aware modeling with comprehensive supervision, balancing efficiency, robustness, and coverage.

\subsection{Online Deployment and Sample Engineering}
\begin{figure}[htbp]
\vspace{-0.2cm}
  \centering
\includegraphics[width=0.9\linewidth]{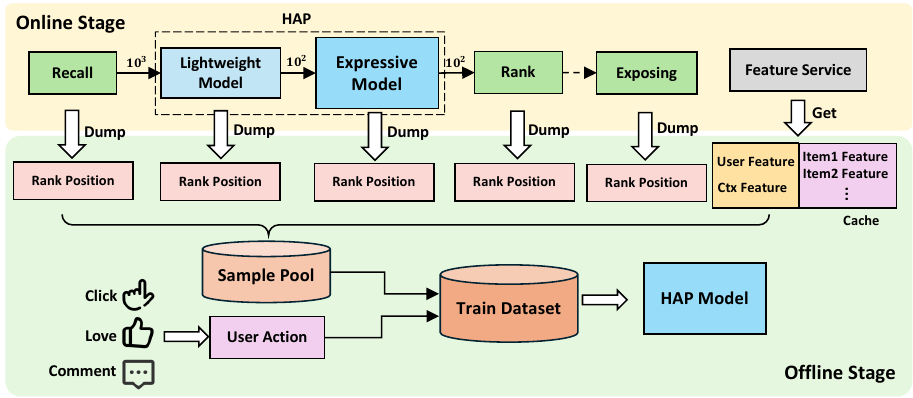}
  \caption{HAP's online deployment in Toutiao: a two-stage pre-ranking pipeline with a lightweight model for easy candidates and an expressive model for hard ones; samples are grouped and common features are shared for efficiency.}
  \label{fig:online}
  \vspace{-0.3cm}
\end{figure}

In this section, we present our practical experience of deploying the HAP framework in the Toutiao recommendation system.
To accommodate HAP in the large-scale online environment, we reconstruct the pre-ranking module into a cascade architecture and build a real-time, list-wise sample pipeline for efficient model training and effect attribution.
The overall system design is illustrated in Fig \ref{fig:online}.

\subsubsection{A Two-Stage Pre-ranking Architecture}
The conventional pre-ranking module in Toutiao employs a single-stage design that sorts thousands of candidates from recall and forwards the top few hundred to the fine-ranking stage.
However, this monolithic structure is incompatible with HAP, which applies different models (lightweight vs. expressive) to candidates of varying difficulty.
To enable HAP and achieve better efficiency–effectiveness trade-offs, we re-architect the system as a two-stage cascade.
In the first stage, a lighweight model with reduced feature dimensions and simplified layers filters out obviously irrelevant candidates, reducing the set size to $G$.
The gating threshold $G$ is determined empirically to balance computational cost and model performance, as detailed in Sec.~\ref{sec:routing_analysis}.
This stage prioritizes inference speed, as excessive complexity yields only marginal benefits compared to its computational cost.
The second stage adopts an expressive model to refine the remaining candidates.
Since this stage naturally focuses on harder cases, we train the model exclusively on hard negatives.
This specialization enhances its discriminative capacity, allowing it to better separate subtle positive–negative distinctions.
Overall, this two-stage implementation of HAP significantly improves ranking precision without introducing noticeable latency overhead.

\subsubsection{Efficient Data Collection and Sample Generation}
To further reduce system load, we optimize the data collection and sampling process with a fully-connected, list-wise sample dumping mechanism.
For each user request, the system logs all user, item, and contextual features, along with the full candidate list.
To minimize redundancy, we apply a shared feature reference strategy in which user-level features are stored only once per request, reducing per-request storage by more than 60\%.
Additionally, we define a 40-minutes feedback window to capture most user interactions.
Delayed behaviors are asynchronously attached afterward to ensure label completeness and long-tail coverage.
This design not only improves data efficiency but also ensures more stable model training by optimizing data storage and reducing redundancy.
\section{EXPERIMENT}
\subsection{ToutiaoRec DataSet}
In recommender systems, data plays a crucial role in evaluating models.
The development of high-performing algorithms heavily relies on the availability of high-quality datasets.
Most open-source datasets \cite{Gao_2022,yuan2022tenrec} focus on specific stages of the recommendation pipeline, lacking full-pipeline observability for single user requests. 
To address this gap, we introduce \textbf{ToutiaoRec}, a large-scale, publicly available industrial dataset designed to support end-to-end optimization in recommender systems.
Table \ref{tab:dataset_comparison} summarizes the similarities and differences between ToutiaoRec and existing datasets.
Unlike them, ToutiaoRec offers full-stage coverage, rank-level annotations, and adheres to strict privacy guarantees.

ToutiaoRec is constructed by sampling 70 million user requests from the Toutiao feed scenario over 3 days, resulting in 313 million requests.
The dataset includes real user feedback such as clicks, likes, comments, and more.
For each request, we retain all exposed items, along with the ranking information across all pipeline stages.
To handle the large volume of unexposed data, we sample negatives as follows:
1) Global random negatives (GN): randomly sampled from the entire candidate pool at a 1:2 ratio to exposed positives.
2) Ranking negatives (RN): 10 unexposed candidates ranked below position 50 in the ranking stage.
3) Pre-ranking negatives (PRN): 10 unexposed candidates ranked below position 200 in the pre-ranking stage.
All types of negatives retain their ranking positions from the respective stages.
In addition, the dataset includes 29 types of anonymized features, such as user behaviors and item attributes.
Given the diverse content formats in Toutiao (articles, microblogs, short videos, etc.), each candidate is annotated with its content type.
To ensure user privacy and legal compliance, we adopt a strict two-step anonymization procedure following GDPR\footnote{The General Data Protection Regulation of the European Union} standards. 
ToutiaoRec is available\footnote{https://github.com/Toutiao-Rec/ToutiaoRec}, with further details provided therein.

\begin{table}[t]
\centering
\caption{Comparisons of ToutiaoRec with existing datasets.}
\label{tab:dataset_comparison}
\small
\begin{tabular}{lccccccc}
\toprule
\textbf{Dataset} & \textbf{Multi-Stage} & \textbf{Complete} & \textbf{Multi-Type} & \textbf{Industrial} \\
 & \textbf{Coverage} & \textbf{Rank Info} & \textbf{Feedback} & \textbf{-Scale} \\
\midrule
MovieLens\cite{harper2015movielens} & \ding{55} & \ding{55} & \ding{55} & \ding{55}  \\
Amazon\cite{ni2019justifying}    & \ding{55} & \ding{55} & \ding{55} & \ding{55} \\
Yelp\cite{asghar2016yelp}      & \ding{55} & \ding{55} & \ding{55} & \ding{55} \\
Taobao\cite{zhu2018learning}    & \ding{55} & \ding{55} & \ding{51} & \ding{51} \\
Tenrec\cite{yuan2022tenrec}    & \ding{55} & \ding{55} & \ding{51} & \ding{51} \\
KuaiRec\cite{Gao_2022}  & \ding{55} & \ding{55} & \ding{51} & \ding{51} \\
\textbf{ToutiaoRec} & \ding{51} & \ding{51} & \ding{51} & \ding{51} \\
\bottomrule
\end{tabular}
\vspace{-0.5cm}
\end{table}

\vspace{-0.2cm}
\subsection{Experiment Setup}
\subsubsection{Sample Construction}
Since existing public datasets lack full-pipeline coverage, they are inadequate for directly validating our method.
We therefore conduct experiments on our publicly released ToutiaoRec dataset.
The data are chronologically split: the first 80\% of requests (by timestamp) are used for training, and the remaining 20\% for evaluation, reflecting real-world production settings.
During training, we leverage all four types of negatives: exposed negatives (EN), ranking negatives (RN), pre-ranking negatives (PRN), and global random negatives (GN).
In evaluation, we simulate the proposed Difficulty-Aware Model Routing (DAMR).
Negatives are categorized by their difficulty and routed accordingly: hard negatives (EN and RN) are scored by the expressive model, while easy negatives (PRN and GN) are handled by the lightweight model.
Positive samples are consistently defined as clicked items (EP) and are included in all test sets by default.
To rigorously assess model performance across negative types, we construct disaggregated test sets listing only the negative samples: TEN (EN), TRN (RN), TPRN (PRN), and TGN (GN).
To further evaluate performance across difficulty levels, we define aggregated sets:
THard (EN + RN) for hard negatives and TEasy (PRN + GN) for easy negatives.
These two grouping strategies correspond to the complementary motivations of analyzing performance by negative type and by difficulty level.
All subsequent experiments follow this evaluation protocol.

\subsubsection{Implementation Details}
We implement HAP in TensorFlow, organizing samples by request with a fixed sequence length of 45 and a global batch size of 1350.
The model is based on DeepFFM: the lightweight model has 3 shallow layers, while the expressive model employs a deeper architecture with attention and genre-specific towers to capture complex patterns.
Both models apply the Gradient-Harmonized Contrastive Learning (GHCL), computing InfoNCE losses per negative type to mitigate gradient conflicts, and a global BCE loss to handle requests without positive samples.
We use the RMSPropV2 optimizer with an initial learning rate of 0.01. 
Training is conducted on 3000 CPU cores for 10 million steps, taking approximately 12 hours.
For reproducibility, code is available\footnote{https://github.com/Toutiao-Rec/HAP}.

\subsubsection{Evaluation Metrics}
\begin{itemize}
\item \textbf{AUC} quantifies the model’s overall ability to distinguish positives from negatives across all users.
\item \textbf{CTR} reflects the ratio of clicked to exposed items, indicating user engagement and overall recommendation attractiveness.
\item \textbf{Duration} measures the average time users stay in the app, reflecting the model’s ability to drive long-term engagement.
\item \textbf{Active Days} indicates the average number of days users remain active, reflecting retention and sustained interest.
\item \textbf{Latency} records the time between receiving a request and returning a response, reflecting system responsiveness. 
\item \textbf{Model Paramters \& FLOPs} 
indicate model complexity in terms of trainable parameters and floating-point operations.
\item \textbf{Cost Ratio} reports the relative change in total cost for offline training and online serving compared with the previous SoTA.

\end{itemize}

\vspace{-0.5em}
\subsubsection{Compared Methods}
We benchmark HAP against several representative and state-of-the-art pre-ranking models.
To ensure a fair comparison, all baseline models are configured to have a comparable parameter size ($\approx$ 6.2M) to our HAP network.
\begin{itemize}
     \item \textbf{DSSM}\cite{huang2013learning}: projects users and items into a low-dimensional semantic space for efficient retrieval via precomputed embeddings.
    \item \textbf{COLD}\cite{wang2020cold} consists of a flexible deep model supporting cross-feature interactions and real-time system optimizations.
    \item \textbf{COPR}\cite{zhao2023copr} employs order-aligned modules and chunk sampling to improve cross-stage ranking consistency.
    \item \textbf{HCCP}\cite{Zhao_2025hccp} integrates upstream and downstream signals to mitigate sample selection bias and improve consistency.
    \item \textbf{HAP} is our heterogeneity-aware adaptive pre-ranking model, composed of a lightweight model (1.8M) for simple candidates and an expressive model (6.2M) for complex ones.
    \item \textbf{$\textbf{HAP}_\text{lite}$} is a cost-efficient variant of HAP that retains competitive effectiveness while reducing computation, with smaller model configurations (1.8M + 4.4M parameters).
 \end{itemize}

\begin{table}[t]
\centering
\caption{Offline AUC comparison of SoTA pre-ranking models on various test sets. The best result is in \textbf{bold} and the second-best result is \underline{underline}, respectively.}
\label{tab:model_performance_sota}
\resizebox{\linewidth}{!}{  
\begin{tabular}{l cccccc}
\toprule
\textbf{Model} & \textbf{TEN} & \textbf{TRN} & \textbf{TPRN} & \textbf{TGN} & \textbf{THard} & \textbf{TEasy} \\
\midrule
DSSM & 0.6901 & 0.8247 & 0.9008 & 0.9785 & 0.7527 & 0.9257 \\
COLD & 0.6936 & 0.8303 & 0.9038 & 0.9803 & 0.7601 & 0.9261 \\
COPR & 0.7221 & 0.8533 & 0.9172 & 0.9871 & 0.7824 & 0.9402 \\
HCCP & 0.7244 & 0.8653 & 0.9218 & \underline{0.9874} & 0.7917 & 0.9441 \\
HAP & \textbf{0.7377} & \textbf{0.8693} & \textbf{0.9236} & \textbf{0.9878} & \textbf{0.8023} & \textbf{0.9468} \\
$\text{HAP}_\text{lite}$ & \underline{0.7331} & \underline{0.8666} & \underline{0.9225} & 0.9872 & \underline{0.7998} & \underline{0.9452} \\
\bottomrule
\end{tabular}
}
\vspace{2pt}
\begin{minipage}{\linewidth}
\footnotesize
\textbf{Model sizes:} DSSM (6.2M), COLD (6.2M), COPR (6.2M), HCCP (6.2M), HAP (1.8M + 6.2M), $\text{HAP}_\text{lite}$ (1.8M + 4.4M).
\end{minipage}
\vspace{-0.9cm}
\end{table}
\vspace{-0.3cm}
\subsection{Offline Performance Comparison}
As shown in Table~\ref{tab:model_performance_sota}, HAP consistently outperforms all state-of-the-art (SoTA) pre-ranking models across both negative types and difficulty levels.
On disaggregated sets (TEN, TRN, TPRN, TGN), improvements are most evident on hard negatives (EN, RN), while easier negatives (PRN, GN) show moderate gains.
For aggregated difficulty sets, performance increases with sample difficulty, with THard exhibiting larger improvements than TEasy.
Although the combined parameter count of HAP’s two-stage models exceeds that of single-stage baselines, the expressive model operates on only a fraction of hard candidates, resulting in lower overall FLOPs and comparable serving latency.
We further report results for \textbf{$\text{HAP}_\text{lite}$}, a cost-efficient variant that retains robustness and strong performance under a smaller budget.
Additional evaluations with alternative metrics yield consistent conclusions (see Table~\ref{tab:model_performance} in appendix).

\vspace{-0.2cm}
\subsection{Ablation Study}
\subsubsection{Ablation Study for GHCL}
\begin{figure}[b]
\vspace{-0.45cm}
  \centering
  \includegraphics[width=0.7\linewidth]{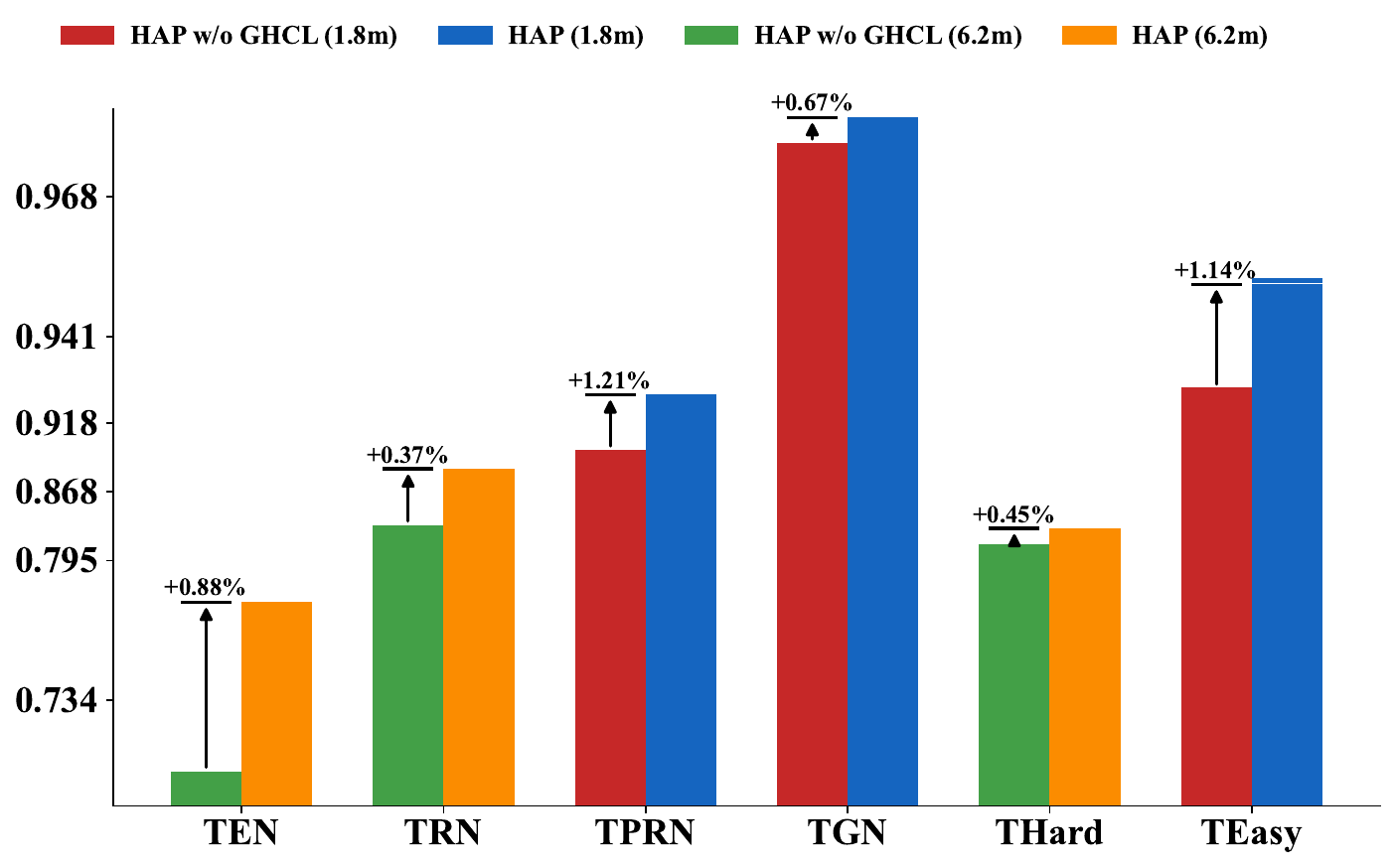}
  \caption{Ablations of GHCL in HAP: removing GHCL leads to a notable AUC decrease across all test sets.}
  \label{fig:ablation_ghcl}
\end{figure}

The Gradient-Harmonized Contrastive Learning (GHCL) module coordinates gradient contributions across negative types, improving training stability and representation learning.
We first remove GHCL from HAP and observe a notable AUC drop across all test sets, as shown in Fig~\ref{fig:ablation_ghcl}, highlighting its importance.
Second, we investigate how GHCL addresses challenges arising from mixing multiple negative types. We compare models trained under three paradigms in Table~\ref{tab:GHCL_general}: models using individual negative types in \texttt{SingleNeg}, the naive mixing strategy (Using subscript \texttt{Mix}) in \texttt{MultiNeg}, and our GHCL. 
We observe three key patterns:
1) Training on a single negative type achieves the best performance on its corresponding test set but induces a seesaw effect, with performance on other negative types dropping significantly.
2) Mixing all negatives without distinction yields more balanced results across types but still underperforms GHCL.
3) GHCL consistently outperforms both \texttt{SingleNeg} and \texttt{MultiNeg}, improving not only individual test sets but also aggregated sets across easy and hard negatives.
Overall, these results validate that GHCL is essential for performance, particularly in multi-negative.

\begin{table}[t]
\centering
\caption{Comparison of negative sampling strategies. For each negative type (GN, PRN, RN, EN),
\texttt{SingleNeg} evaluates BCE loss versus the combined InfoNCE+BCE objective. \texttt{MultiNeg} combines all four. Best results in \textbf{bold} and \underline{underline}.}
\label{tab:GHCL_general}
\begin{minipage}{1\linewidth} 
\centering
\resizebox{\linewidth}{!}{
\begin{tabular}{l p{2.1cm}  cccccc}
\toprule
\textbf{NegType} & \textbf{Model} & \textbf{TEN} & \textbf{TRN} & \textbf{TPRN} & \textbf{TGN} & \textbf{THard} & \textbf{TEasy} \\
\midrule
\multirow{8}{*}{SingleNeg} 
 & $\text{BCE}_\text{GN}$ & 0.5930 & 0.6859 & 0.7717 & \textbf{\underline{0.9992}} & 0.6391 & 0.8680 \\
 & $\text{BCE}_\text{PRN}$ & 0.6186 & 0.7732 & \textbf{\underline{0.9616}} & 0.9406 & 0.6955 & \textbf{\underline{0.9502}} \\
 & $\text{BCE}_\text{RN}$ & 0.6238 & \textbf{\underline{0.9017}} & 0.8595 & 0.8861 & 0.7623 & 0.8728 \\
 & $\text{BCE}_\text{EN}$ & \textbf{\underline{0.7591}} & 0.6815 & 0.7121 & 0.7815 & 0.7207 & 0.7484 \\
 & $\text{InfoNCE}_\text{GN}$ & 0.5941 & 0.6834 & 0.7677 & 0.9991 & 0.6378 & 0.8634 \\
 & $\text{InfoNCE}_\text{PRN}$ & 0.6148 & 0.7564 & 0.9531 & 0.9348 & 0.6845 & 0.9487 \\
 & $\text{InfoNCE}_\text{RN}$ & 0.6417 & 0.8973 & 0.8556 & 0.8749 & \textbf{\underline{0.7677}} & 0.8618 \\
 & $\text{InfoNCE}_\text{EN}$ & 0.7577 & 0.6851 & 0.7167 & 0.7788 & 0.7213 & 0.7357 \\
\midrule
\multirow{3}{*}{MultiNeg} 
 & $\text{BCE}_\text{Mix}$ & 0.7232 & 0.8626 & 0.9173 & 0.9826 & 0.7875 & 0.9371 \\
 & $\text{InfoNCE}_\text{Mix}$ & 0.7234 & 0.8634 & 0.9233 & 0.9872 & 0.7882 & 0.9430 \\
 & GHCL & \textbf{\underline{0.7261}} & \textbf{\underline{0.8662}} & \textbf{\underline{0.9295}} & \textbf{\underline{0.9898}} & \textbf{\underline{0.7938}} & \textbf{\underline{0.9488}} \\
\bottomrule
\end{tabular}
}
\end{minipage}
\vspace{-0.65cm}
\end{table}

\begin{table*}[htbp]
\centering
\begin{threeparttable} 
\caption{Ablations of DAMR in HAP: Online A/B test results on Toutiao, trained on the large-scale logs.}
\label{tab:online_cost}
\begin{tabular}{l cccccccc}
\toprule
\textbf{Model} & 
\textbf{THard\tnote{a}} & \textbf{TEasy\tnote{a}} & \textbf{CTR\tnote{c}} &
\textbf{Compute Cost\tnote{b}} & 
\textbf{FLOPs ($10^6$)} &
\textbf{Latency ($ms$)} &
\textbf{Cost Ratio\tnote{c}} \\
\midrule
Unified Large Model & 0.8158 & 0.9586 & +0.55\% & 38.1 & 70 & 225 & +15\% & \\
 HAP & 0.8197 & 0.9579 &+3.22\% & 39.2 & 89 & 203 & -6\% & \\
\bottomrule
\vspace{-10pt}
\end{tabular}
\begin{tablenotes}
\footnotesize
\item[] \textsuperscript{a} Evaluation on Full-scale Industrial Data; 
\textsuperscript{b} Training Duration $\times$ CPU kcores (h·kcores);
\textsuperscript{c} Relative gain over the previous SoTA.
\end{tablenotes}
\end{threeparttable}
\end{table*}

\subsubsection{Ablation Study for DAMR}
We further investigate the role of DAMR in balancing model effectiveness and efficiency.
As detailed in Table~\ref{tab:model_comparison} in the appendix, model performance consistently improves with capacity, revealing a clear trade-off between accuracy and computational cost.
While a unified large model yields notable gains on hard samples, it incurs redundant computation on easy ones.
In contrast, HAP employs DAMR to route only hard candidates to the expressive model, maintaining accuracy while improving efficiency.
To isolate the effect of DAMR, we compare HAP with a \emph{Unified Large Model}, which shares the same architecture as HAP’s expressive component but operates in a single-stage manner.
As shown in Table~\ref{tab:online_cost}, although HAP has slightly more parameters, its serving costs drops by about 6\%, yielding comparable or even lower latency.
Notably, on the THard dataset, HAP even outperforms the \emph{Unified Large Model}.
This stems from DAMR's specialization: the expressive model is trained and served exclusively on hard negatives, mitigating optimization conflicts and enabling expert-level discrimination.
These advantages are further validated in the online environment, where HAP delivers higher CTR and lower latency than the \emph{Unified Large Model}.
This consistent improvement across both offline and online settings underscores DAMR’s effectiveness in achieving a superior balance between performance and deployability in large-scale recommender systems.

\vspace{-0.2cm}
\subsection{In-depth Analysis of HAP Components}
\label{sec:analysis}
\subsubsection{Analysis of Gradient Harmonization}
\begin{figure}[t]
  \centering
  \begin{subfigure}[b]{\linewidth}
    \centering
    \includegraphics[width=0.9\linewidth]{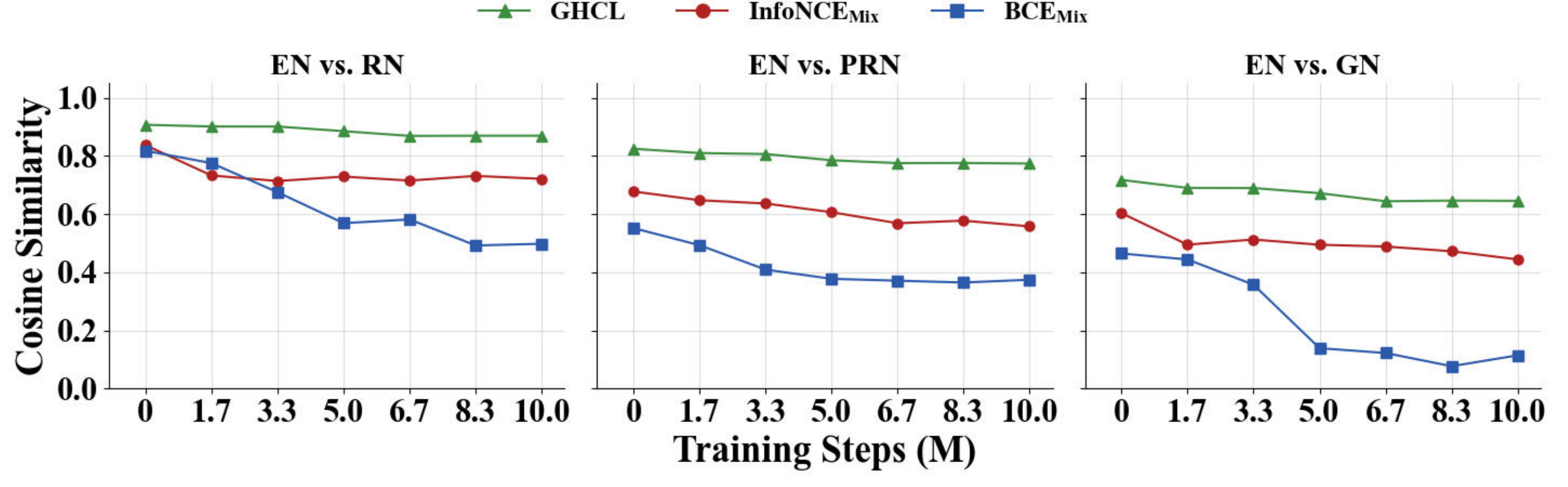}
    \caption{Inter-set gradient cosine similarities.}
    \label{fig:cos_fig}
  \end{subfigure}

  \vspace{0.5em}

  \begin{subfigure}[b]{\linewidth}
    \centering
    \includegraphics[width=0.9\linewidth]{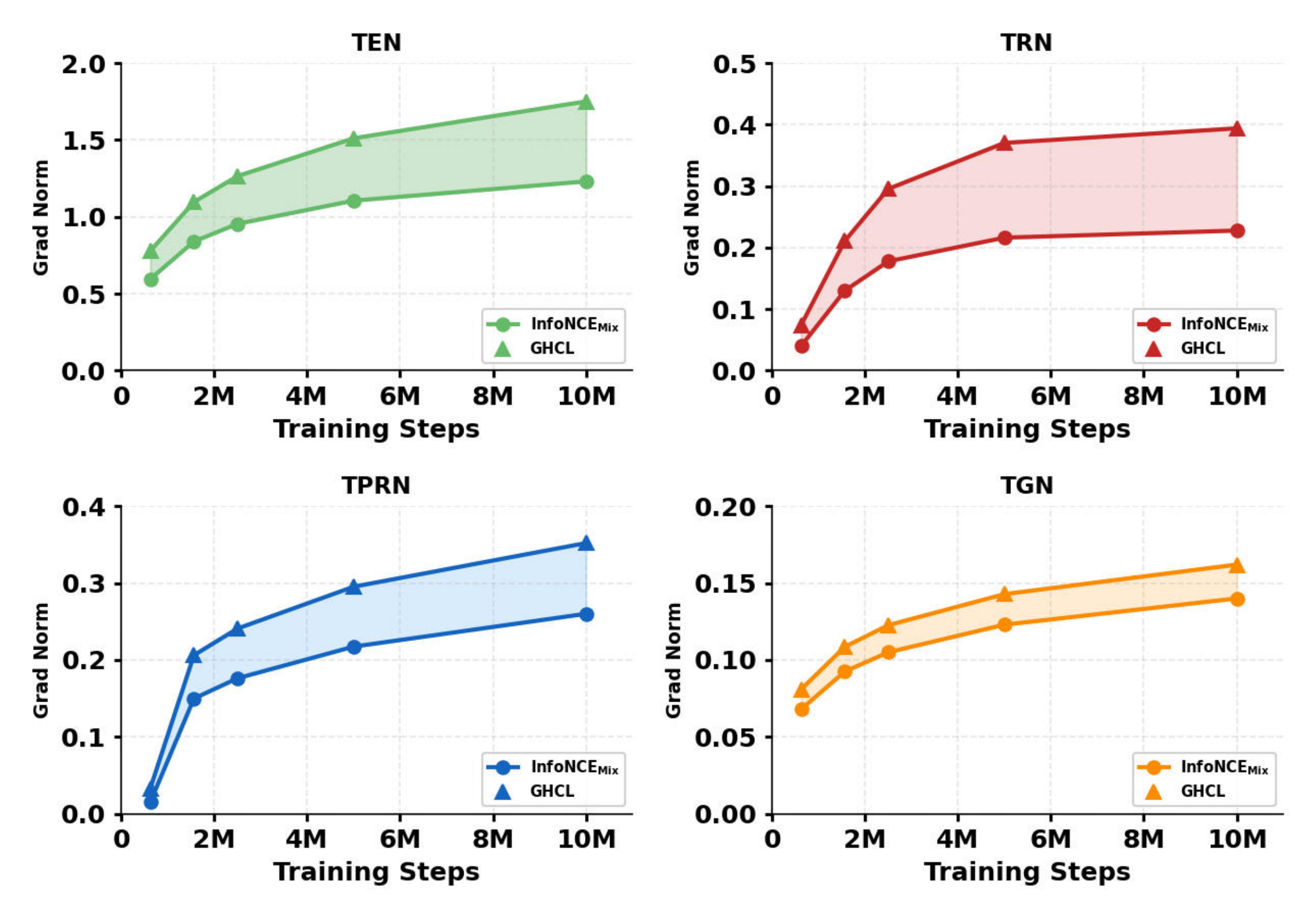}
    \caption{Intra-set gradient norms.}
    \label{fig:grad_norm}
  \end{subfigure}
  \caption{Gradient statistics of heterogeneous negatives.
  GHCL stabilizes directions and magnitudes, mitigating conflicts compared to existing methods.
  }
  \label{fig:grad_analysis}
  \vspace{-0.6cm}
\end{figure}

We conduct a gradient-level analysis to empirically verify GHCL's effectiveness in mitigating gradient conflicts across samples of varying difficulty.
Specifically, we track both 1) the cosine similarity between gradients from different negative sets and 2) the gradient norms within each set.
As illustrated in Fig~\ref{fig:grad_analysis}, GHCL maintains higher inter-set cosine values, indicating more consistent gradient directions, and stabilizes the gradient magnitudes across easy and hard samples.
In contrast, $\text{BCE}_\text{Mix}$ and $\text{InfoNCE}_\text{Mix}$ exhibit increasing angular divergence and imbalance in gradient strength as training progresses.
These findings confirm that GHCL effectively harmonizes gradient optimization, preventing destructive interference and promoting stable convergence across heterogeneous negatives.

\subsubsection{Routing Strategy Analysis}
\label{sec:routing_analysis}
\begin{figure}[t]
  \vspace{-0.1cm}
  \centering
  \includegraphics[width=0.9\linewidth]{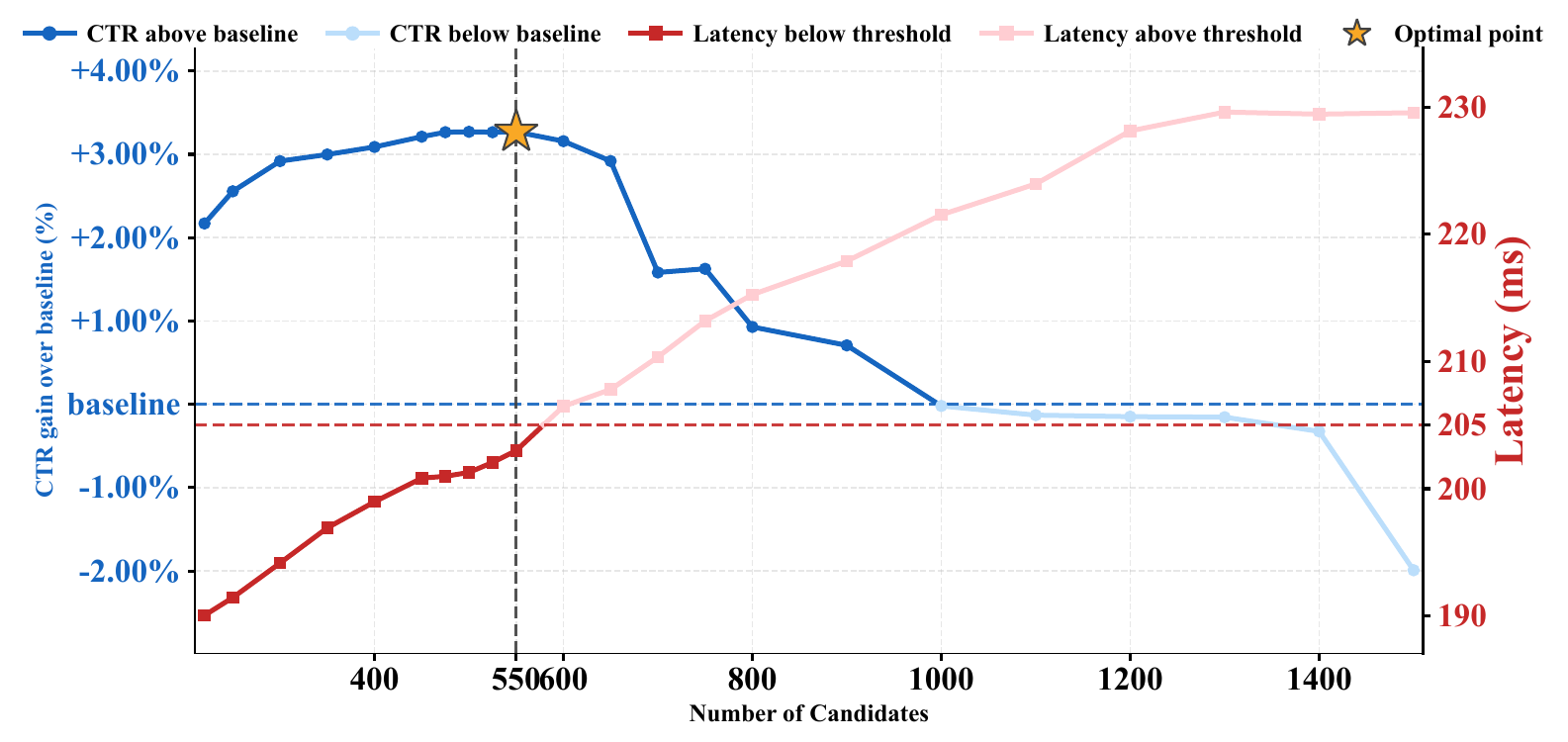}
  \caption{Online performance–efficiency trade-off of HAP under varying gating thresholds. }
  \label{fig:ctrVSlatency}
  \vspace{-0.6cm}
\end{figure}

We further analyze the routing strategy that determines how many candidates are forwarded from the lightweight to the expressive model in the two-stage serving system.
While offline evaluations are performed on static test sets, online deployment is more dynamic and sensitive to latency constraints.
The number of routed candidates critically affects both serving cost (latency) and model effectiveness.
To quantify this trade-off, we conduct online A/B tests by varying the gate threshold $G$, which determines the number of candidates evaluated by the expressive model.
As shown in Fig~\ref{fig:ctrVSlatency}, CTR initially increases with higher routing ratios but drops once system load exceeds capacity, mainly due to request failures and the fact that the expressive model is trained exclusively on hard samples.
This trade-off curve provides a principled way to select the optimal gating threshold under given serving constraints in large-scale industrial systems.

\subsection{Online Deployment and Performance}
To further evaluate the real-world applicability of HAP, we deployed it in the production environment of Toutiao.
The online evaluation used large-scale recommendation logs that contains user interactions and full request traces across all stages, including retrieval, pre-ranking, ranking, and re-ranking, collected over 60 days.
Two models were trained: a lightweight model (3.1M) and an expressive model (8.6M).
Based on our routing analysis, the gating threshold was set to 550, meaning only the top 30\% of candidates from the lightweight model were re-ranked by the expressive one.
We conducted an online A/B test against the previous production SoTA model.
Results show that HAP achieved \textbf{+0.05\%} gain in user active days, \textbf{+0.4\%} in app duration, and \textbf{+3.0\%} in CTR. 
Considering that Toutiao serves hundreds of millions of users, even a 0.01\% increase in user activity is practically meaningful, making the observed 0.05\% improvement particularly noteworthy.
Moreover, HAP maintained comparable serving latency to the SoTA while reducing CPU usage by 6\%, demonstrating its ability to deliver higher effectiveness at lower computational cost in a large-scale production environment.
Following there positive outcomes, HAP has been fully deployed across the Toutiao production system for the past 9 months.

\section{Conclusion}
In this paper, we present HAP, a novel framework for tackling candidate heterogeneity in pre-ranking.
Unlike conventional approaches constrained by gradient conflicts and computational inefficiency, HAP integrates 
Gradient-Harmonized Contrastive Learning (GHCL) for stable optimization and Difficulty-Aware Model Routing (DAMR) for adaptive resource allocation.
Extensive experiments on our newly released public dataset and large-scale online deployment in Toutiao demonstrate that HAP consistently improves recommendation performance while reducing serving cost.
These results establish HAP as a practical and scalable solution for building more effective and efficient industrial recommender systems.

\bibliographystyle{ACM-Reference-Format}
\balance
\bibliography{sample-base}

\clearpage
\appendix

\section{Appendix}

\subsection{Model Complexity and Effectiveness Trade-off Analysis}

As shown in Table \ref{tab:model_comparison}, model performance improves with increased model complexity, reflecting a trade-off between effectiveness and efficiency.
However, experiments also show that a unified large model improves more significantly on hard samples than on easy ones as complexity grows.
In contrast, HAP uses DAMR to allocate computation precisely to the "hard candidate", avoiding wasteful computation on the "easy candidate". This leads to higher efficiency without sacrificing performance.

\begingroup
\setlength{\intextsep}{4pt}          
\setlength{\textfloatsep}{4pt}       
\setlength{\abovecaptionskip}{2pt}   
\setlength{\belowcaptionskip}{2pt}   

\begin{table}[!h] 
\centering
\caption{Performance and Efficiency Trade-offs of Models with Different Complexities}
\label{tab:model_comparison}

\begin{subtable}[htbp]{\columnwidth}
  \centering
  \footnotesize
  \setlength{\tabcolsep}{3.5pt}
  \captionsetup{aboveskip=2pt, belowskip=2pt}
  \caption{Performance Comparison}
  \begin{tabular}{llcccccc}
    \toprule
    \textbf{Metric} & \textbf{Model} & \textbf{TEN} & \textbf{TRN} & \textbf{TPRN} & \textbf{TGN} & \textbf{THard} & \textbf{TEasy} \\
    \midrule
    AUC & Light  & 0.7192 & 0.8617 & 0.9267 & 0.9886 & 0.7836 & 0.9471 \\
    AUC & Medium & 0.7219 & 0.8625 & 0.9270 & 0.9887 & 0.7867 & 0.9475 \\
    AUC & Heavy  & 0.7261 & 0.8662 & 0.9295 & 0.9898 & 0.7938 & 0.9488 \\
    \bottomrule
  \end{tabular}
\end{subtable}

\vspace{2pt} 

\begin{subtable}[htbp]{\columnwidth}
  \centering
  \footnotesize
  \begin{threeparttable}
  \setlength{\tabcolsep}{4pt}
  \captionsetup{aboveskip=2pt, belowskip=2pt}
  \caption{Efficiency Comparison}
  \begin{tabular}{lccc}
    \toprule
    \textbf{Model} & \textbf{Compute Cost\tnote{a}} & \textbf{Params} & \textbf{FLOPs ($10^6$)} \\
    \midrule
    Light  & 28.5 & 1.8m & 32 \\
    Medium & 33.9 & 4.4m & 47 \\
    Heavy  & 36.1 & 6.2m & 57 \\
    \bottomrule
  \end{tabular}
  \begin{tablenotes}
    \footnotesize
    \item[a] Training Duration $\times$ CPU kcores (h·kcores)
  \end{tablenotes}
  \end{threeparttable}
\end{subtable}

\end{table}
\endgroup

\vspace{-10pt} 

\subsection{Evaluation on Multiple Offline Metrics}

To provide a more complete comparison, we evaluated our models using additional metrics besides AUC. Specifically, we introduced Group AUC (GAUC) and Logloss. The results are shown in Table~\ref{tab:model_performance}. We can see that the performance trends on GAUC and Logloss are perfectly consistent with the AUC results. This indicates that the improvements from our HAP model are robust and significant.

\begin{table}[b]
\centering
\caption{Comparative Evaluation of HAP Against Industrial Recommender Baselines. 
The best results are highlighted in bold and \underline{underline}. Metrics with $\uparrow$ indicate higher is better, while $\downarrow$ indicates lower is better.}
\label{tab:model_performance}

\resizebox{\columnwidth}{!}{%
\begin{threeparttable}
\begin{tabular}{lccccccc}
\toprule
 & \textbf{Model} & \textbf{TEN} & \textbf{TRN} & \textbf{TPRN} & \textbf{TGN} & \textbf{THard} & \textbf{TEasy} \\
\midrule
\multirow{6}{*}{AUC$\uparrow$} 
 & DSSM & 0.6901 & 0.8247 & 0.9008 & 0.9785 & 0.7527 & 0.9257 \\
 & COLD & 0.6936 & 0.8303 & 0.9038 & 0.9803 & 0.7601 & 0.9261 \\
 & COPR & 0.7221 & 0.8533 & 0.9172 & 0.9871 & 0.7824 & 0.9402 \\
 & HCCP & 0.7244 & 0.8653 & 0.9218 & \underline{0.9874} & 0.7917 & 0.9441 \\
 & HAP & \textbf{0.7377} & \textbf{0.8693} & \textbf{0.9236} & \textbf{0.9878} & \textbf{0.8023} & \textbf{0.9468} \\
 & $\text{HAP}_\text{lite}$ & \underline{0.7331} & \underline{0.8666} & \underline{0.9225} & 0.9872 & \underline{0.7998} & \underline{0.9452} \\
\midrule
\multirow{6}{*}{GAUC$\uparrow$\tnote{a}} 
 & DSSM  & 0.6483 & 0.8229 & 0.9002 & 0.9794 & 0.7377 & 0.9277 \\
 & COLD  & 0.6499 & 0.8245 & 0.9021 & 0.9809 & 0.7393 & 0.9337 \\
 & COPR  & 0.6617 & 0.8389 & 0.9112 & 0.9875 & 0.7514 & 0.9404 \\
 & HCCP & 0.6659 & 0.8545 & 0.9184 & \underline{0.9891} & 0.7602 & 0.9449 \\
 & HAP  & \textbf{0.6783} & \textbf{0.8627} & \textbf{0.9203} & \textbf{0.9897} & \textbf{0.7691} & \textbf{0.9473} \\
 & $\text{HAP}_\text{lite}$ & \underline{0.6736} & \underline{0.8565} & \underline{0.9194} & 0.9887 & \underline{0.7664} & \underline{0.9457} \\
 \midrule
\multirow{6}{*}{LogLoss$\downarrow$\tnote{b}} 
 & DSSM & 0.0489 & 0.0422 & 0.0381 & 0.0325 & 0.0453 & 0.0353 \\
 & COLD & 0.0482 & 0.0420 & 0.0378 & 0.0323 & 0.0451 & 0.0351 \\
 & COPR & 0.0473 & 0.0409 & 0.0372 & 0.0322 & 0.0447 & 0.0341 \\
 & HCCP & 0.0471 & 0.0403 & 0.0366 & \underline{0.0319} & 0.0438 & 0.0337 \\
 & HAP & \textbf{0.0459} & \textbf{0.0385} & \textbf{0.0359} & \textbf{0.0315} & \textbf{0.0434} & \textbf{0.0332} \\
 & $\text{HAP}_\text{lite}$ & \underline{0.0461} & \underline{0.0398} & \underline{0.0362} & 0.0321 & \underline{0.0435} & \underline{0.0335} \\
\bottomrule
\end{tabular}
\begin{tablenotes}
\footnotesize
\item[a] GAUC computes AUC at the user level, reflecting personalized ranking performance within each user's context.
\item[b] LogLoss quantifies the gap between predicted probabilities and ground truth. Lower values indicate better accuracy.
\end{tablenotes}
\end{threeparttable}%
}
\end{table}

\end{document}